\documentclass[12pt]{article}

\def\la{\mathrel{\mathpalette\fun <}}
\def\ga{\mathrel{\mathpalette\fun >}}
\def\fun#1#2{\lower3.6pt\vbox{\baselineskip0pt\lineskip.9pt
\ialign{$\mathsurround=0pt#1\hfil##\hfil$\crcr#2\crcr\sim\crcr}}}

\begin{document}
\newcommand{\qt}{\tilde q}
\newcommand{\Tr}{{\rm Tr}\,}
\newcommand{\nfour}{\mbox{${\cal N}\!\!=\!4\;$}}
\newcommand{\ntwo}{\mbox{${\cal N}\!\!=\!2\;$}}
\newcommand{\none}{\mbox{${\cal N}\!\!=\!1\;$}}
\newcommand{\qtu}{\tilde q_{1}}
\newcommand{\qtd}{\tilde q_{2}}
\newcommand{\md}{\mu_{D}}
\newcommand{\beq}{\begin{equation}}
\newcommand{\eeq}{\end{equation}}
\newcommand{\xv}{\vec{\xi}}
\newcommand{\ax}{\vert\xi\vert}
\newcommand{\xva}{\vert\vec{\xi}\vert}
\newcommand{\rt}{\tilde r}
\newcommand{\ar}{i\theta}
\newcommand{\as}{\alpha^{2}}
\newcommand{\gs}{g^{2}}
\newcommand{\fl}{\phi_{l}}
\newcommand{\fh}{\phi_{h}}
\newcommand{\pu}{\psi_{1}}
\newcommand{\pd}{\psi_{2}}
\newcommand{\vp}{\varphi}
\newcommand{\pt}{\partial}

\renewcommand{\theequation}{\thesection.\arabic{equation}}

\begin{titlepage}
\renewcommand{\thefootnote}{\fnsymbol{footnote}}

\begin{flushright}
PNPI-TH-2512/03\\
ITEP-TH-19/03\\
hep-th/0303047\\

\end{flushright}

\vfil

\begin{center}
\baselineskip20pt
{\bf \Large  Flux Tubes  on  Higgs Branches
in SUSY Gauge Theories }
\end{center}
\vfil

\begin{center}
{\large   K. Evlampiev$^a$ and   A. Yung$^{b,c}$}

\vspace{0.3cm}

$^a${\it Petersburg State University, St. Petersburg, 199034 }\\
$^b${\it Petersburg Nuclear Physics Institute, Gatchina, St. Petersburg,
188300}\\
$^c${\it Institute of   Theoretical and Experimental Physics,
Moscow, 117250}

\vfil

{\large\bf Abstract} \vspace*{.25cm}
\end{center}

We study flux tubes on Higgs branches with curved geometry
in supersymmetric gauge theories.
As a first example we consider \none QED with
one flavor of charges and with Higgs branch
curved by adding a Fayet-Iliopoulos (FI) term. We show that in a
generic vacuum on the Higgs branch  flux tubes exist but become
``thick''.  Their internal structure in the plane orthogonal to the
string is determined by ``BPS core'' formed by heavy fields and
long range ``tail '' associated with light fields living on the
Higgs branch. The string tension is given by  the  tension
of ``BPS core'' plus contribution coming from the ``tail''.
Next we consider \ntwo QCD with gauge group $SU(2)$
and $N_f=2$ flavors of fundamental matter (quarks) with the same mass.
We perturb this theory by the mass term for the adjoint field
which to the leading order in perturbation parameter do not break \ntwo
supersymmetry and reduces to FI term. The Higgs branch has Eguchi-Hanson
geometry.
We work out string solution in the generic vacuum on the Higgs branch
and calculate its string  tension. We also  discuss if
these strings can turn into semilocal strings, the possibility
related to the confinement/deconfinement phase transition.

\vfil

\end{titlepage}

\newpage

\section{Introduction}

The mechanism of
  confinement
 as a dual Meissner effect arising upon condensation of
monopoles was suggested a long ago \cite{MH}.
 Once monopoles condense the electric flux is confined
in the (dual) Abrikosov--Nielsen--Olesen (ANO) flux tube \cite{A,NO}
connecting heavy trial charge and anti-charge. The flux tube  has
constant energy per unit length (string tension ). This
ensures that the confining potential between heavy charge and
anti-charge increases linearly with their separation. However,
because the dynamics of monopoles is hard to control in
strong coupling  gauge theories
no quantitative  description of this phenomenon was constructed in QCD.

The revival of interest to this problem occurs after the work of
 Seiberg and
Witten  \cite{SW1,SW2}. They showed that in supersymmetric gauge theories
the holomorphy and electromagnetic duality are powerful tools
to study   non-Abelian dynamics at strong coupling.
 Using the exact solution of $N=2$
supersymmetric gauge theory they were able to demonstrate
 that the condensation of
monopoles do really occurs near the monopole vacuum
 once $N=2$ theory is slightly perturbed
  by the mass term of the adjoint matter \cite{SW1}.

Since then a lot of papers study confinement and formation
of flux tubes in supersymmetric gauge theories
\cite{DS,HSZ,FG,EFMG,Y99,VY,Y,KS,KnB,ADPU,PU}.
One important particular feature of
supersymmetric theories is the presence of modular spaces -- manifolds
on which scalar fields can develop arbitrary VEV's. If the gauge group
is broken down completely by scalar  VEV's  such vacuum manifolds are called
Higgs branches. In this paper we study the formation of flux tubes
in supersymmetric gauge theories with Higgs branches.

In fact, this problem was addressed earlier in ref. \cite{Y99} for the
case of Higgs branches with flat geometry. In particularly,
 the formation of flux tubes
on the Higgs branch of  \ntwo QCD with gauge group $SU(2)$
 and $N_f=2$ flavors
of fundamental matter hypermultiplets (quarks) with the same mass was studied.
The Higgs branch in this case represent an extreme type I superconductor
with vanishing Higgs mass. It was shown in \cite{Y99} that the flux tubes
still exist in this set up although becomes logarithmically ``thick''
due to the presence of light scalar field. Because of this the confining
potential
behaves as
\beq
\label{cpot}
V(L)\sim \frac{L}{\log{L}},
\eeq
with the separation $L$ between two heavy  well-separated (magnetic) charges.
 This potential  is still confining but is no longer linear in $L$.
Note, that in this case we have  condensation
of scalar components of quarks (electric charges) so these are
monopoles which are confined. As the potential between heavy trial
charges is an order parameter to distinguish  between different
phases of the theory, we see that we have a new confining phase with a non-linear
potential which is specific for Higgs branches in supersymmetric
gauge theories.

In this paper we continue to study flux tubes on Higgs branches
concentrating on the case of Higgs branches with curved geometry.
The curvature on a Higgs branch is induced by adding of the
Fayet-Iliopoulos (FI) term to the  theory.

Our first example is \none QED with one flavor of quarks and the FI D-term
added. This theory has a two dimensional Higgs branch
of a hyperbolic  form.
  In order to regularize this theory in the infrared one
would like to lift the Higgs branch making scalars which fluctuate along
the vacuum manifold slightly massive. It turns out that this can be done without
destroying of \none supersymmetry. We start with \ntwo QED
with one quark flavor which has no Higgs branch. Then we  add a mass term
for the neutral matter field breaking \ntwo supersymmetry  down to \none.
Then in the limit of mass parameter of this field $\mu\to\infty$ we
recover \none QED with its Higgs branch. At large but finite $\mu$
we have Higgs branch lifted but the potential along the would be
Higgs branch becoming more and more flat as we increase $\mu$.

In this set up we study the formation of flux tubes in a generic vacuum
on the would be Higgs branch. We show that flux tubes exist and have
the following structure in the plane orthogonal to the string axis.
They are formed by ``BPS core'' which size is determined by the FI-parameter
and long range ``tail'' formed by light scalar fields.
The string tension is given by  the  tension
of ``BPS core'' plus contribution coming from the ``tail''.
If the VEV of scalar fields squared is of the same order
as the FI-parameter the contribution of the ''tail'' becomes
a small correction to the tension coming from the ''BPS core''.
That is why we can call this string ``almost BPS''.
Note however, that this  string  is not a BPS state and belongs to a long
supermultiplet.

This string gives rise to the confinement of monopoles
with the potential which depends on the monopole separation $L$ as
\beq
\label{cpotcurved}
V(L)\sim L\left(1+\frac{const}{\log{L}}\right).
\eeq
Here the first term comes from the ``BPS core'' while the second one
is determined by the ``tail''.

Next we consider \ntwo QCD with gauge group $SU(2)$ and two flavors of
quarks with the same mass. This theory has four dimensional
Higgs branch \cite{SW2}.
We perturb this theory by the mass term for the adjoint field
which to the leading order in perturbation parameter do not break \ntwo
supersymmetry and reduces to FI term
\cite{HSZ,VY}. The non-zero FI-parameter induces
curvature on  the Higgs branch. It becomes a hyper-Kahler manifold
with  Eguchi-Hanson \cite{EH} geometry.

We work out the ANO string solution in a generic point on the
Higgs branch and
calculate the string tension. Similar to the \none case it is
given by the contribution of ``BPS core''  plus correction coming
from the ``tail'' of light fields living on the Higgs branch.

We also discuss if this ANO string can turn into semilocal string
( see \cite{AV} for a review on semilocal strings)
by increasing the size of its ``BPS core''. If this happen
this would
ruin  confinement making the potential between
 monopoles to fall-off as a power of $L$ at large $L$.
We show that if VEV of the quark fields is much larger then the FI-parameter
the semilocal string is not developed. ANO string appears
to be stable so  monopoles are  confined
by the potential (\ref{cpotcurved}).

However, for vacua at the base $S_2$ cycle of the
Eguchi-Hanson manifold for which quark VEV is equal to the FI-parameter
ANO strings turn into semilocal strings \cite{ADPU,PU}. For these vacua
we have deconfinement phase.  We put forward a conjecture that
confinement/deconfinement phase transition occurs exactly at the
base cycle.  Finally we consider effects which break \ntwo
supersymmetry
down to \none and show that in this case the deconfinement phase
disappears.

The paper is organized as follows. In Sect. 2 we review
flux tubes  on Higgs branches with flat geometry \cite{Y99}.
In Sect. 3 we introduce \none QED with FI term and
in Sect. 4 study flux tubes on the Higgs branch in this theory.
In Sect. 5 we consider \ntwo QCD with gauge group $SU(2)$
and two flavors of quarks with the same mass and
study flux tubes on the Higgs branch in this theory.
In Sect. 6 we discuss the issue of semilocal strings.
Sect. 7 contains our brief conclusions.

\section{Extreme type I strings}

In this section we   review classical solution for
 ANO vortices in theories with flat Higgs potential which
arises in supersymmetric settings \cite{Y99}.
In particular, in ref.\cite{Y99} flux tubes
on the Higgs branch of  \ntwo QCD
with two flavors of fundamental matter (quarks) were studied.
Consider  the
 Abelian Higgs model
\begin{equation}
\label{ah}
S_{AH}=\int d^4x\left\{\frac1{4g^2}\,F^2_{\mu\nu}+|\nabla_\mu
q|^2+\lambda(|q|^2-v^2)^2\right\},
\end{equation}
for the single complex field $q$  with  quartic coupling $\lambda=0$.
Here $\nabla_{\mu}=\partial_{\mu}-in_eA_{\mu}$, where $n_e$ is
the electric charge of the field $q$.
Following \cite{Y99} we consider first this model  with small
$\lambda$,
Then  we take the limit $\lambda\to0$.

The field $q$ develops VEV, $q=v$, breaking down the U(1) gauge
 group.
Photon acquires the mass
\beq
\label{mgamma}
m^2_{\gamma}=2n_e^2g^2v^2,
\eeq
while the Higgs  mass is equal to
\beq
\label{Hmass}
m_q^2=4\lambda v^2.
\eeq

The model (\ref{ah}) is the standard Abelian-Higgs model which admits
ANO strings \cite{A,NO}.
For an arbitrary $\lambda$ the Higgs mass
differs from that of the photon. The ratio of the photon mass to the
 Higgs mass is an
important parameter, in the theory of superconductivity it characterizes
 the type of
superconductor. Namely, for $m_q<m_\gamma$ we have the type I
superconductor in which two well separated ANO strings attract each
other.  Instead for $m_q>m_\gamma$ we have the type II one in which two
well separated strings repel each other. This is related to the fact
that scalar field produces an attraction for two vortices,
while the electromagnetic field produces a repulsion.

Now we consider the extreme type I limit in which
\beq
\label{gammahiggs}
 m_q\ll m_\gamma .
\eeq
We also assume the week coupling regime in the model  (\ref{ah})
$\lambda\ll g^2 \ll 1$.

 The general idea to find the string solution
is to separate behavior of
different  fields at different scales present in the problem due
to the condition (\ref{gammahiggs}). This method goes back to the paper
by Abrikosov \cite{A} in which the tension of type II string
under condition $m_{q}\gg m_{\gamma}$ has been calculated.
The similar idea was used in \cite{Y99} to calculate the tension
of the type I string under condition $m_{q}\ll m_{\gamma}$.

To the leading order in $\log m_\gamma/m_q$ the vortex
solution has the following structure in
the plane orthogonal to the string axis \cite{Y99}. The
electromagnetic field is confined in a core with the radius
\begin{equation}
\label{rad}
R_g\ \sim\ \frac1{m_\gamma}\log{\frac{m_\gamma}{m_q}}\ .
\end{equation}
The scalar field is close to zero inside the core. Instead,
outside the core, the electromagnetic field is vanishingly
small while the scalar field behavies as
\beq
\label{scsol}
q=v\left(1-\frac{K_0(m_q r)}{\log{1/m_q R_g}}\right)
e^{i\alpha},
\eeq
where $r$ and $\alpha$ are  polar coordinates
in the plane orthogonal to he string axis.
Here $K_0$ is the Bessel function
with exponential fall-off at infinity and logarithmic
behavior at small arguments, $K_0(x)\sim \log{1/x}$ at $x\to 0$.
The reason for this behavior is that in the absence of the
electromagnetic field outside the core the scalar field satisfies a free
equation  of motion and (\ref{scsol}) is a solution to this
equation.
From (\ref{scsol})
 we see that the scalar field  slowly (logarithmically)
approaches its boundary value $v$.

 The result for the string tension is \cite{Y99}
\begin{equation}
\label{rten}
T \ =\ \frac{2\pi v^2}{\log{ m_\gamma/m_q}}\ .
\end{equation}
The main contribution to the tension in (\ref{rten})
 comes from the logarithmic ``tail'' of the scalar $q$.
It is given by the kinetic term for the scalar field in (\ref{ah}).
This term contains a logarithmic integral over $r$. Other terms
in the action are suppressed by powers of $\log m_\gamma/m_q$
as compared with the one in (\ref{rten}).

The results in (\ref{rad}), (\ref{rten}) mean that if we naively take the
limit $m_q\to0$ the string becomes infinitely thick and its
tension goes to zero \cite{Y99}. This means that there are no
strings in the limit $m_q=0$.  The absence of ANO
strings in theories with flat Higgs potential was first noticed in
\cite{PRTT}.

One might think that the
 absence of ANO strings means that there is no
  confinement of monopoles in  theories
with  Higgs branches. As
we will see now this is not  the case \cite{Y99}.
 So far we have considered infinitely long ANO strings. However
the setup for the confinement problem is slightly different.
 We have to consider monopole--anti-monopole pair at
large but finite separation $L$. Our aim is to take the limit
$m_q\to0$. To do so let us consider ANO string of the finite
length $L$ within the region
\begin{equation}
\label{L}
\frac1{m_\gamma}\ \ll\ L\ \ll\ \frac1{m_q}\ .
\end{equation}
Then it turns out that $1/L$ plays the role of the $IR$-cutoff
 in Eqs. (\ref{rad}) and (\ref{rten}) instead of $m_q$ \cite{Y99}.
The reason for this is that for $r\ll L$ the problem is two dimensional
and the solution of the  two dimensional free
 equation of motion for a scalar field is
given by (\ref{scsol}). If we naively put $m_q=0$ in this solution
the Bessel function reduces to the logarithmic function which cannot
reach a finite boundary value at infinity. Thus as we mentioned
above the infinitely long flux tubes do not exist.  This  was noticed
 in \cite{PRTT}. However for $r\gg L$ the problem becomes three
dimensional. The solution for the three dimensional free scalar
equation of motion behaves as $q-v \sim 1/|x|$,
where $x_n$, $n=1,2,3$ is a coordinate in the three dimensional
space. We see that with this behavior the scalar field  reaches its
boundary value at infinity. Clearly $1/L$ plays a role of IR
cutoff for the logarithmic behavior of the scalar field.

 Now we can safely put $m_q=0$.
The result for the electromagnetic core of the vortex becomes
\begin{equation}
\label{ts}
R_g\ \sim\ \frac1{m_\gamma}\log{m_\gamma L}\ ,
\end{equation}
while its string tension is given by \cite{Y99}
\begin{equation}
\label{ct}
T\ =\ \frac{2\pi v^2}{\log{ m_\gamma L}}\ .
\end{equation}

We see that the ANO string becomes "thick" but still its
transverse size $R_g$ is much less than its length $L$, $R_g\ll
L$. As a result the potential between heavy well separated
monopole and anti-monopole is still  confining but is
no longer linear in $L$. It behaves
as \cite{Y99}
\begin{equation}
V(L)\ =\ 2\pi v^2\, \frac L{\log{ m_\gamma L}}\ .
\end{equation}
 As we already explained in the Introduction the
potential $V(L)$ is an order parameter which distinguishes
different phases of a theory
(see, for example, review
\cite{IS}). We conclude that we have a new confining phase on the
Higgs branches. It is clear that
this phase can arise only in supersymmetric theories
because we do not have Higgs branches without supersymmetry.

\section{\none QED with Fayet-Iliopoulos term}
\setcounter{equation}{0}

\subsection{The model}

The field content of \none QED is as follows. The vector multiplet
contains U(1) gauge field $A_{\mu}$ and Weyl fermion
$\lambda^{\alpha}$, $\alpha=1,2$. The chiral matter multiplet contains
two complex scalar fields $q$ and $\qt$ as well as two complex Weyl
fermions $\psi^{\alpha}$ and $\tilde{\psi}_{\alpha}$. The bosonic part
of the action reads
\beq
\label{qed}
S_{QED} =\int d^4 x \left\{ \frac1{4 g^2}F_{\mu\nu}^2 +
\bar{\nabla}_{\mu}\bar{q}\nabla_{\mu}q+
\bar{\nabla}_{\mu}\tilde{q}\nabla_{\mu}\bar{\tilde{q}}
+V(q,\qt)\right\},
\eeq
where $\nabla_{\mu}=\partial_{\mu} -\frac{i}{2}A_{\mu}$,
$\bar{\nabla}_{\mu}=\partial_{\mu} +\frac{i}{2}A_{\mu}$, so
we assume that that matter fields have electric charge $n_e=1/2$.
The potential of this theory comes from the $D$-term and
given by
\beq
\label{pot1}
V(q,\qt)=\frac{g^2}{8}\left( |q|^2 -|\tilde{q}|^2
-\xi_3\right)^2,
\eeq
where parameter $\xi_3$ arises if we include FI $D$-term in
our theory. We denote here the FI $D$-term parameter $\xi_3$.
This notation will become clear later in \ntwo setup
where $D$ and $F$ FI parameters form a triplet with respect to
global $SU(2)_R$.

The vacuum manifold of the
theory (\ref{qed}) is a Higgs branch determined
by the condition
\beq
\label{1hb}
\vert{<q>}\vert^{2} - \vert{<\qt>}\vert^{2}
          = \xi_{3}.
\eeq
The dimension of this Higgs branch is two. To see this note, that we
have two complex scalars (four real variables) subject to
one constraint (\ref{1hb}). Also we have to subtract one gauge
phase, thus we have 4-1-1=2.

Now let us consider the mass spectrum of the QED (\ref{qed}).
As it is clear from (\ref{1hb}) at any non-zero $\xi_3$ scalar fields
develop VEV's breaking $U(1)$ gauge symmetry. The photon mass
is given by
\beq
\label{mgamma1}
m^2_{\gamma}=\frac12g^2v^2,
\eeq
where we introduce the VEV of scalar field
\beq
\label{xi}
v^2=\vert{<q>}\vert^{2} + \vert{<\qt>}\vert^{2}.
\eeq

To find matter masses we diagonalize the $4\times 4$ mass matrix for
scalar fields in (\ref{pot1}). It has three zero eigenvalues one
of which is "eaten" by the Higgs mechanism while two others correspond
to chiral massless multiplet living on the Higgs branch. The remaining
fourth eigenvalue is equal to the photon mass (\ref{mgamma1}),
\beq
\label{heavy}
m_H=m_{\gamma}.
\eeq
This eigenvalue corresponds to  the scalar superpartner
 of the photon in the massive \none vector multiplet.

Our aim is to study string solutions in a generic vacuum
on the Higgs branch (\ref{1hb}). It is clear that this solution is
formed by both massive electromagnetic and scalar fields as well as
by massless scalars living on the Higgs branch. Similar to the approach
of ref. \cite{Y99} (reviewed above in Sect. 2) we would like to
regularize the problem at hand in the infrared giving these massless
scalars a small mass $m_L$ and then taking the limit $m_L\to 0$.
We will do it in the next subsection.

\subsection{ Softly broken \ntwo QED }

It turns out that we can lift the Higgs branch of \none QED
considered in the previous subsection giving the massless fields small
masses without breaking \none supersymmetry.  To do so
following ref. \cite{VY} let us consider  \ntwo QED.
This theory besides fields which enter \none QED contains also
neutral chiral field  $A$ which interacts with charged matter via
superpotential
\beq
\label{sup}
W_{N=2}=\frac1{\sqrt{2}}\tilde{Q}AQ-\frac1{2\sqrt{2}}\eta\, A ,
\eeq
where $Q$ and $\tilde{Q}$ are charge superfields and
 we also introduce the FI $F$-term proportional to the complex
FI parameter
\beq
\label{eta}
\eta=\xi_1+i\xi_2 .
\eeq
Two FI $F$-term parameters $\xi_{1,2}$ together with one $D$-term
parameter $\xi_3$ form a triplet with respect to global $SU(2)_R$
symmetry of \ntwo QED.

Let us break \ntwo supersymmetry down to \none adding a mass term
for the chiral field $A$ via additional superpotential
\beq
\label{brsup}
\delta W  =\frac{\mu}{2}A^2.
\eeq
Now if we take the limit $\mu\to\infty$ we can integrate out heavy
$A$-field and end up with \none QED considered in the  previous
subsection. Namely, adding (\ref{brsup}) to (\ref{sup})    and
integrating out $A$ we get the superpotential
\beq
\label{supint}
W=-\frac1{4\mu}\left(\tilde{Q}Q-\frac{\eta}{2}\right)^2.
\eeq

 This superpotential leads to the following
scalar potential in \none QED (\ref{qed}) \cite{VY}
\beq
\label{pot}
V(q,\qt)={g^{2} \over 8}(\vert{q}\vert^{2} - \vert{\qt}\vert^{2}
          - \xi_{3})^{2}+ \frac1{4 \mu^{2}}(\vert{q}\vert^{2} +
          \vert{\qt}\vert^{2}){\bigl\vert}q\qt -
 {\eta \over 2}{\bigr\vert}^{2}
\eeq
The first term here comes from the D-components of the gauge multiplet,
 while the second one comes from the superpotential above.
We expand $\eta=\xi_1+i\xi_2$, where $\xi_1$ and $\xi_2$ are real and use
$SU(2)_R$ rotation to put $\xi_2=0$.

We see that in the limit $\mu\to\infty$ this potential reduces to the
one in (\ref{pot1}). However, for any finite $\mu$ the potential
(\ref{pot}) has no Higgs branch and the vacuum state is uniquely
defined. Namely,
the potential (\ref{pot}) has a minimum at
$$
<q>=\sqrt{\xi_3}\cosh{\rho_0},
$$
\beq
\label{vac}
<\qt>=\sqrt{\xi_3}\sinh{\rho_0},
\eeq
where we introduce parameter $\rho_0$ defined via
\beq
\label{rho0}
\sinh{2\rho_0}=\frac{\xi_1}{\xi_3}
\eeq
Note, that if $\xi_2=0$ we can use gauge freedom to
make both $<q>$ and $<\tilde{q}>$ positive.

Calculating the $4\times4$ scalar mass matrix near this vacuum we
obtain one zero eigenvalue (corresponding to the  state "eaten" by
the Higgs mechanism) and another one equal to
the mass of the photon, see (\ref{mgamma1}), (\ref{heavy}).
As we explained above the corresponding scalar together with the
photon form a bosonic part of \none vector massive multiplet.
The mass of this multiplet
 remains unchanged by our IR regularization. The only
modification is that the VEV of the scalar field is now fixed by the
parameters of the theory. Namely, it is defined by (\ref{xi}) which can
be now expressed in terms of FI parameters as
\beq
\label{xidef}
v^4=\xi^2_1+\xi^2_3.
\eeq

The   other two
eigenvalues of the scalar mass matrix are given by
\beq
\label{ml}
m_L=\frac{\xi_1}{\mu} .
\eeq
 This is the mass of one
\none chiral multiplet (containing two real scalars).
In the limit of large $\mu$ that we consider here
\beq
\label{con}
m_L\ll m_{\gamma}.
\eeq
This means that we have one heavy scalar in our problem ( with the mass
equal to the mass of the photon, $m_H=m_{\gamma}$
) and two light scalars. In particularly,
in the limit of $\mu\to\infty$
 $m_L$ goes to zero and we recover the Higgs branch of \none QED.

Note that the form of the second term in the potential (\ref{pot})
is not important for our purposes. It serves as a IR regularization
which lifts the Higgs branch and gives would be massless moduli fields
a small mass (\ref{ml}).

\section{String solution}
\setcounter{equation}{0}

In this section we consider solutions for flux tubes which
arise in a generic point on the Higgs branch of \none QED.
First we start with a base point on the Higgs branch,
which corresponds to $<\tilde{q}>=0$. Next we consider
the generic point on the vacuum manifold.

\subsection{BPS strings}

Consider the particular vacuum with  $<\tilde{q}>=0$ in
\none QED (\ref{qed}).
It is well known that
there is a BPS ANO string solution for this particular
choice of vacuum \cite{GS}.

We can easily recover this solution in the softly broken QED with the
scalar potential (\ref{pot}) taking the limit $\xi_1=0$, while
keeping $v=\xi_3$ non-zero.   It is clear that light fields do not
 play any role  and we can look for
 the string  solution using the following {\em ansatz}
 \beq
\label{bpsan}
\qt=0 .
\eeq
With this substitution the bosonic part of softly broken \ntwo QED
reduces to the standard Abelian Higgs model
(\ref{ah}) for
one complex scalar field $q$ interacting with the electromagnetic field,
at the particular
value of the quartic coupling $\lambda=g^2/8$. This value of $\lambda$
ensures that the mass of the scalar field in (\ref{ah}) is equal to the
mass of the photon, see  (\ref{heavy}). As we already explained in
the previous section this is a consequence of \none supersymmetry.

Thus our model is on the
 boundary separating superconductors of
the  I and II type.
In this case vortices do not interact.
It is well known that vanishing of the interaction at $m_H=m_\gamma$ can be
explained  by the BPS nature of the ANO strings.  The ANO string
satisfy the first order equations and saturate the Bogomolny bound~\cite{B}.
 This
bound follows from the following representation  for the string tension $T$,
\begin{equation}
T =2\pi\xi_3 \,n+\int \!{\rm d}^2
x\left\{\left[\frac1{2g}F_{ik}+\frac{g}{4}
\left(|q|^2-\xi_3\right)\varepsilon_{ik}\right]^2
+ \frac12 \left|\nabla_i \,q+i\varepsilon_{ik}
\nabla_k \, q\right|^2 \right\}.
\label{tens}
\end{equation}
Here indices $i,k=1,2$ denote coordinates transverse to the axis of the
vortex.
The minimal value of the tension is reached when  both  positive
terms in the integrand of Eq.~(\ref{tens}) vanish. The string tension becomes
\beq
 T_{BPS}\ =\ 2\pi\xi_3 \, n \ ,
\label{tension}
\eeq
where the winding number $n$ counts the magnetic flux $2\pi\,n$ (we assume
positive $n$).  The linear dependence of string tensions
 on $n$ implies the absence of
their interactions.

For simplicity we consider the case $n=1$.
Putting the integrand in Eq.~(\ref{tens}) to zero gives
 two first order differential equations,
$$
r\frac{\rm d}{{\rm d}r}\,\phi (r)- f(r)\,\phi (r)\ =\ 0\ ,
$$
\beq
 -\frac1r\,\frac{\rm d}{{\rm d}r} f(r)+\frac{g^2}{4}\,
\left(\phi^2(r)-\xi_3\right)\ =\ 0\ ,
\label{foe}
\eeq
where the profile functions $\phi(r)$ and $f(r)$ are introduced
in a standard way,
\begin{eqnarray}
\label{prof}
q(x) &=& \phi (r)\, {\rm e}^{i\,\alpha}\ ,\nonumber\\
A_i(x) &=&\! \!-2\epsilon_{ij}\,\frac{x_j}{r^2}\ [1-f(r)]\ .
\end{eqnarray}
Here $r=\sqrt{x^2_i}$ is the distance and $\alpha$ is the polar angle in the
(1,2) plane. The  profile functions are real and
satisfy the boundary conditions
\begin{eqnarray}
&& \phi (0)=0\ ,
\qquad ~f(0)=1\ , \nonumber\\ && \phi (\infty)=\sqrt{\xi_3}\ , \quad
f(\infty)=0\ ,
\label{bc}
\end{eqnarray}
which ensures that the scalar field reaches its vev $\sqrt{\xi_3}$ at the
infinity and the vortex carries one unit of the magnetic flux.
The equations (\ref{foe}) with boundary conditions (\ref{bc})
lead to the unique solution for the profile functions
(although an analytic form of this solution is not found).

In QED with the \ntwo supersymmetry broken down to \none the emergence
of the first order equations (\ref{foe}) signals that some (half)   of
the remaining four SUSY charges of \none algebra act trivially on the
ANO solution (cf. \cite{HS,DDT,GS,VY}). In this case the Bogomolny
(topological) bound for the string tension coincides with the central
charge of SUSY algebra. Note that at $\mu=0$ we have BPS strings
in \ntwo QED \cite{HSZ,VY}. As we increase $\mu$ but keep $\xi_1=0$ they
become BPS strings of \none QED. As the number of states in
string multiplets cannot jump we conclude that we get two
BPS string
multiplets at  large $\mu$ from one BPS string in \ntwo theory at
$\mu=0$.

\subsection{String in a generic vacuum}

Now let us turn to the generic case when both $\xi_1$ and $\xi_3$ are non-zero.
It is clear that in this case the ANO string is no longer BPS saturated.
To see this, note that now we cannot take the light scalar fields
with mass $m_L$ to be zero on the vortex solution. This means that
the vortex has a long range ``tail'' in $(1,2)$ plane
formed by the light scalars. This
ensures attraction of different vortices and type I superconductivity.


The  string tension for non-BPS string is bounded from below by the
central charge of the \none algebra
\beq
\label{sb}
T\ge 2\pi \xi_3,
\eeq
which is determined by the FI D-term parameter $\xi_3$ \cite{GS}.

Now let us work out the approximate solution for the ANO vortex and
calculate its tension.
The difference of the
problem at hand with the one studied in \cite{Y99}
and reviewed in Sect. 2 is that in
\cite{Y99}  was considered the case
which corresponds to $\xi_3=0$  in the low energy QED. In this case
heavy scalar does not develop VEV and can be put to zero on
the vortex solution. The only scalar which is present in the problem is
the light one.

 In the present case, when both  $\xi_1$ and $\xi_3$ are non-zero
both heavy and light scalars are non-zero.
To solve the problem
 we adopt the following model for the
 structure of the vortex in $(1,2)$ plane. The electromagnetic field
 together with
 the heavy scalar form a core of  relatively small
radius. Let us call it  $R_c$.
Outside this core  heavy fields are almost zero, while   light
components   produce a ``tail' of size $1/m_L$.
 To be more specific we consider separately three different
regions: $r\la R_c$, $R_c \ll r \ll 1/m_L$ and $r \ga 1/m_L$.

In the first region (inside the core)
\beq
\label{reg1}
r\la R_c
\eeq
only heavy scalar and electromagnetic fields are non-zero.
while the light scalars are almost zero.
  This suggests that we can look for the vortex solution with
\beq
\label{tq}
\qt = 0.
\eeq
With this ansatz our model reduces to the standard Abelian Higgs model
(\ref{ah})
 at the special value of quartic coupling $\lambda=g^2/8$, see
Sect 4.1.
This means that the solution for the vortex inside the core
is given by the BPS solution we reviewed in the
previous subsection. In particular, the
field $q$  and $A_{i}$ are given by (\ref{prof}) with the
profile functions $\phi$ and $f$ subject to the first order equations
(\ref{foe}) and boundary conditions (\ref{bc}). Now the boundary
condition (\ref{bc}) at infinity should be understood as a condition at
the boundary of the core , namely, at $r\ga R_c$.
Say, for scalar fields this means
\beq
\label{1bc}
q(r\ga R_c)=\sqrt{\xi_3},
\eeq
while $\qt$ is zero. This means that outside the core
scalar fields go to the base point on the vacuum manifold.

As soon as the size of this BPS string is given by
$1/\sqrt{g^2\xi_3}$ we conclude that
\beq
\label{Rc}
R_c= \frac1{\sqrt{g^2\xi_3}}.
\eeq

The string tension of the  vortex with $n=1$ is given by
\beq
\label{rept}
T=2\pi \xi_3 + T_{tail}.
\eeq
Here the first term is ``BPS'' contribution (\ref{tension}) coming from
the region (\ref{reg1}) inside the core, while $T_{tail}$ stands for
the contribution of the light scalar ``tail'' coming from the
region outside the core. Let us work out this contribution.

Consider the region of intermediate $r$ outside the core,
\beq
\label{reg2}
R_c \ll r \ll 1/m_L.
\eeq
In this region we  can neglect the second term in the potential
(\ref{pot}).
The motion of scalar fields  is restricted by the constraint
\beq
\label{const}
\vert{q}\vert^{2} - \vert{\qt}\vert^{2}
          = \xi_{3}
\eeq
enforced by the first term in the potential (\ref{pot}), see also
 (\ref{pot1}). The constraint (\ref{const})
ensures that the only light scalar modes (actually massless, once
we neglect $m_L$) are exited in the region (\ref{reg2}). To put it another
way, the constraint (\ref{const}) reduces the motion of scalar
fields to the motion along the Higgs branch given by (\ref{const}).
From (\ref{const}) we see that this Higgs branch has non-flat metric
(see also eq.(\ref{sigma}) below). This makes a difference with the case
studied in \cite{Y99} where Higgs branch was flat.

Now let us work out the sigma model on the Higgs branch.
The constraint (\ref{const}) can be solved by the substitution
$$
q=\sqrt{\xi_3}\,e^{i\alpha+i\beta(x)}\cosh \rho(r),
$$
\beq
\label{rho}
\bar{\qt}=\sqrt{\xi_3}\,e^{i\alpha-i\beta(x)}\sinh \rho(r).
\eeq

Two complex fields $q$ and $\qt$ have two different phases. We use the
gauge freedom to fix their average to be equal to the polar
angle $\alpha$ in order to ensure the correct flux of the solution.
The phase difference contains arbitrary function $\beta(x)$.
The  boundary condition  for functions $\rho$ and $\beta$
is fixed by VEV's of scalar fields (\ref{vac}).
$$
\rho(r\sim R_c)= 0,
$$
\beq
\label{brbc}
\rho(r\sim 1/m_L)=\rho_0,\; \beta(r\sim 1/m_L)=0.
\eeq

Now let us take  the low energy limit formally sending
$m_{\gamma}\to\infty$. In this limit we can integrate out the gauge fied
$$
A_i=-i\frac{\bar{q}\partial_i q - \partial_i \bar{q}q
+\qt\partial_i \bar{\qt} - \partial_i \qt\bar{\qt}}
{ \bar{q} q  + \qt \bar{\qt} }
$$
\beq
\label{gaugepot}
 =2\left(\partial_i \alpha +\frac{\partial_i \beta}{\cosh{2\rho}}
\right),
\eeq
where we use the substitution (\ref{rho}). Then our  model
 reduces to the 2d sigma model
\beq
\label{sigma}
T_{tail}= \xi_3\int d^2 x  \cosh 2\rho\,\left\{ (\partial_{i}
\rho )^2 + (\partial_{i}\beta)^2\tanh^2{2\rho}\right\}
\eeq
with non-flat metric of the target space.

Now the problem is to find the classical solution
for two dimensional sigma model (\ref{sigma}) with
boundary conditions (\ref{brbc}). To do this in more general
setting we consider a sigma model with arbitrary metric
\beq
\label{sigmagen}
T_{tail}= \xi_3 \int d^2 x\, g_{MN}
\partial_{i}\vp^N  \partial_{i}\vp^N,
\eeq
where $N,M$ numerates light scalar fields. Now we
 assume that these fields depend only on the radial coordinate
$r$. This leads us to the one dimensional sigma model
which determine the tension of the string "tail"
\beq
\label{sigma1d}
T_{tail}= \frac{2\pi\xi_3}{\log{1/m_L R_c}}\int_0^1 d t\,   g_{MN}
\partial_{t}\vp^N  \partial_{t}\vp^N,
\eeq
where we introduce normalized logarithmic time
\beq
\label{t}
t=\frac{\log{r/ R_c}}{\log{1/m_L R_c}}.
\eeq

The equations of motion for this sigma model
define a geodesic line
\beq
\label{geod}
\partial^2_t\vp^N+\Gamma^N_{MK}\partial_t\vp^M \partial_t\vp^K =0,
\eeq
where $\Gamma^N_{MK}$ denotes the connection on the target manifold.
The energy conservation shows that the action on this
line is determined by the square of the length of this line
\beq
\label{tail}
T_{tail}= \frac{2\pi\xi_3}{\log{1/m_L R_c}}\,l^2.
\eeq
Here  the length of the geodesic line reads
\beq
\label{length}
l=
\int_0^1 d t \sqrt{g_{MN}
\partial_{t}\vp^N  \partial_{t}\vp^N}.
\eeq

For our model (\ref{sigma}) at hand the geodesic line is particularly
simple. Clearly,
\beq
\beta=0
\eeq
at the geodesic line
and its length on the Higgs branch becomes
\beq
\label{lengthhip}
l=
\int_0^{\rho_0} d \rho  \sqrt{\cosh{2\rho} },
\eeq
where the upper limit is
defined by (\ref{rho0}). This length determine our final
result for the tension of the  string
\beq
\label{ten}
T= 2\pi\xi_3 + \frac{2\pi\xi_3}{\log{(g\sqrt{\xi_3}/m_L )}}\,l^2.
\eeq
It is easy to check that the third region $r\ga1/m_L$
(where scalar fields approach their VEV's exponentially) gives
corrections to this result suppressed by powers of
$\log{g\sqrt{\xi_3}/m_L }$, see also \cite{Y99}.

Let us note that this string solution "feels"  not only
the VEV of the scalar field $v^2$ but the whole structure of the
Higgs branch. In particular, the size of the core $R_c$ is
determined by $g\sqrt{\xi_3}$ (see (\ref{Rc})) rather then
by the mass of the photon (\ref{mgamma1}) ( determined by $gv$).

Of course, if we send $\xi_1$ to zero  going to the base point on the
Higgs branch $<\qt>=0$
 the second term in (\ref{ten}) vanishes and we will get
the BPS string. Note however, that the long non-BPS string multiplet
contains two bosonic and two fermionic states,
while the short BPS multiplet contains
one bosonic and one fermionic state. As the number of states
cannot jump the long multiplet
 reduces to two short BPS multiplets at $\xi_1=0$.
This is in accordance with our conclusion in the end of the
previous subsection where we considered breaking of \ntwo
supersymmetry by turning on $\mu$. There we saw that
one \ntwo BPS multiplet reduces to two  \none BPS multiplets
as we switch on $\mu$ at zero $\xi_1$.

If $\xi_3$ and $v^2$ are of the same order  the second term
in (\ref{ten}) becomes small as compared with the first one
due to the logarithmic suppression. This is the reason why we
can call
this string "almost BPS". Note however, that this string is
not a BPS one. It does not belong to a short BPS multiplet
and all four supercharges act on this solution non-trivially.

To make contact with the case discussed in Sect. 2
let us consider the limit $\xi_3\ll v^2$. In this limit
the geometry of the Higgs branch becomes flat and
$l^2\to v^2/\xi_3$. Then our result (\ref{ten}) reduces
to eq.  (\ref{rten}). Note that as we send $\xi_3$ to
zero the size of the core of the string $R_c$ grows and
eventually freezes as it reaches the value $R_g$, see (\ref{rad}).
At $g\sqrt{\xi_3}\ll 1/R_g$ the core is determined by the
electromagnetic field \cite{Y99} and the heavy scalar plays no role
any longer.

Let us now remove our IR regularization sending $\mu\to\infty$. Then
we recover the Higgs branch of \none QED and light scalars become
strictly massless. Repeating the arguments in Sect. 2 we see that
infinitely long strings do not exist any
longer  in the generic point on the
Higgs branch. However, strings of finite length still exist. Their
string tension is now given by
\beq
\label{tenL}
T= 2\pi\xi_3 + \frac{2\pi\xi_3}{\log{(g\sqrt{\xi_3}\,L)}}\,l^2.
\eeq
They give rise to the confinement of monopoles with the potential
of type (\ref{cpotcurved}).

To conclude this section let us make a comment on literature.
In ref. \cite{ADPU,PU} vortices in \none QED were considered and
it was  concluded that in generic point on the Higgs branch
the string is unstable.  The only vacuum which support  string
solutions is the base point of the Higgs branch $<\qt>=0$. The so called
"vacuum selection rule" was put forward in \cite{ADPU,PU}
 to ensure this property.
We would like to stress that our results here and in ref. \cite{Y99}
do not contradict above mentioned papers. As we already explained
infinitely long strings do not exist in a generic point on the modular
space, indeed. However this does not mean that we loose confinement.
As we explained in Sect. 2 to study confinement we have to consider
strings of finite length. Strings of finite length do exist
in a generic point on the Higgs branch and produce confining
potential (\ref{cpotcurved}).

\section{\ntwo QCD }
\setcounter{equation}{0}

In this section we study another example of flux tubes on
Higgs branches. We consider \ntwo QCD with gauge group $SU(2)$
and two flavors of fundamental matter (quarks). If masses of these
quarks are equal then there is a Higgs branch in this theory.
First we review  the effective low energy theory on the
Higgs branch \cite{SW2} and then construct the string  solution.

\subsection{Higgs branch}

The \ntwo vector multiplet of the  theory at hand
on the component level consists of  the
gauge field $A^a_\mu$, two Weyl fermions $\lambda^{\alpha a}_1$
and $\lambda^{\alpha a}_2$ $(\alpha=1,2)$ and the complex scalar
$\varphi^a$, where $a=1,2,3$ is the color index. Fermions form
a doublet  $\lambda^{\alpha a}_f$ with respect to global $SU(2)_R$
group, $f=1,2$.

The scalar potential of this theory has a flat direction.
The adjoint scalar field  develop an arbitrary VEV along this
direction breaking $SU(2)$ gauge group down to $U(1)$. We choose
$\langle\varphi^a\rangle=\delta^{a3}\langle a\rangle$. The
complex parameter $\langle a\rangle$ parameterize the
Coulomb branch. The low energy effective
theory  generically contains only the photon $A_\mu=A^3_\mu$ and its
superpartners: two Weyl fermions $\lambda^{3\alpha}_f$
and the complex scalar $a$. This is massless short vector \ntwo
multiplet. It contains 4 boson + 4 fermion states.
  W-boson and its superpartners are massive with masses
 of order of $\langle a\rangle$.

Quark hypermultiplets have the following structure. They
consist of complex scalars $q^{kA}$,   $\tilde{q}_{Ak}$
and fermions $\psi^{k\alpha A}$,
$\tilde \psi^{\alpha}_{Ak}$, where  $k=1,2$
is the color index and $A=1,\ldots,N_F$ is the flavor one.
Scalars $q^{kA}$,   $\bar{\tilde{q}}^{kA}$
form a doublet $q^{kfA}$, $f=1,2$ with respect to $SU(2)_R$ group.
All these states are in the BPS short representations of
\ntwo algebra  on the Coulomb
branch with $4\times N_c\times  N_f=16$ real boson states (+ 16
fermion states).

Coulomb branch has three singular points where
monopoles , dyons or charges become massless. Two
of them correspond to monopole and dyon singularities of the
pure gauge theory. Their positions on the Coulomb branch are
given by \cite{SW2}
\begin{equation}
\label{mds}
u_{m,d}\ =\ \pm\ 2m\Lambda_2-\frac12\Lambda^2_2\ ,
\end{equation}
where $u=\frac12\langle\varphi^{a^2}\rangle$ and $\Lambda_2$ is
the scale of the theory with $N_f=2$.
In the large $m$ limit $u_{m,d}$ are
approximately given by their values in the pure gauge theory
$u_{m,d}\simeq\pm2m\Lambda_2=\pm2\Lambda^2$, where $\Lambda$
is the scale of $N_f=0$ theory.

The charge singularity corresponds to the point where half of quark
states
becomes massless. We denote them $q^{fA}$ and  $\psi^{\alpha A}$,
$\tilde \psi^{\alpha}_{A}$ dropping the color index. They form
$N_f=2$ short hypermultiplets with $4\times  N_f=8$ real boson states.
The rest of  quark states acquire large mass $2m$ and we ignore them
in the low energy description.
The charge singularity appears  at the point
\begin{equation}
\label{acs}
a\ =\ -\ \sqrt2\ m\
\end{equation}
on the Coulomb branch.
 In terms of variable $u$ (\ref{acs}) reads
\begin{equation}
\label{cs}
u_c\ =\ m^2+\frac12\Lambda_2^2.
\eeq
Strictly speaking, we have $2+N_f=4$ singularities on the
Coulomb branch. However, two of them  coincides for the case of
two flavors of matter with the same mass.

The effective theory on the Coulomb branch near the charge
singularity (\ref{acs}) is given by
${\cal N}=2$ QED with light matter fields
$q^{fA}$, and their superpartners  as well as the
photon multiplet.

We deform the underlying non-Abelian theory adding a superpotential
\beq
\label{ubr}
\delta W=\frac{\mu}2 \, \Phi^{a2} ,
\eeq
which is a mass term for the adjoint chiral field
$\Phi^a$. Generally speaking
this perturbation breaks \ntwo supersymmetry down to \none.
The Coulomb branch shrinks to three above mentioned singular
points which we call \none vacua. In particularly, here we
will be interested in quark vacuum (\ref{cs}) which is far away
from the origin in week coupling provided the mass of quarks is
large $m\gg \Lambda_2$.

In the low energy effective theory near quark vacuum the
superpotential (\ref{ubr}) can be expanded as
\beq
\label{ffi}
\delta W=-\frac1{2\sqrt{2}}\eta\delta A +\cdots,
\eeq
where $A$ is the neutral chiral superfield  with the  lowest component
$a$, while
\beq
\label{ximu}
\eta=-2\sqrt{2}<a>=4\mu\,m.
\eeq
Dots in (\ref{ffi}) denote higher orders in $\delta A=A+\sqrt{2}m$.

It turns out that if in the limit of small $\mu$ we truncate
the series in (\ref{ffi}) restricting ourselves only to the
leading in $\delta A$ term then  the perturbation (\ref{ffi}) does not
break \ntwo supersymmetry in the low energy QED \cite{HSZ,VY}.  The
reason for this is that the leading term in (\ref{ffi}) is linear in
$\delta A$. It is FI $F$-term which does not break \ntwo supersymmetry.

The bosonic part of the effective \ntwo QED near quark vacuum
looks like
$$
S^{QED}=\int d^4x\left\{\frac1{4g^2}
F^2_{\mu\nu}+ \frac1{g^2}|\partial_{\mu} a|^2+
\bar\nabla_\mu\bar q_{Af}\nabla_\mu q^{fA}+
\right.
$$
\begin{equation}
\label{rtact}
\left.
\frac{g^2}8[\mbox{ Tr }(\bar q\tau_m q)-\xi_m]^2\right\}
+\frac12 |q^{fA}|^2|a+\sqrt{2}m_A|^2,
\end{equation}
where trace is calculated over flavor and $SU(2)_R$ indices
while $\xi_m$, $m=1,2,3$ is a $SU(2)_R$ triplet of FI parameters.
In particular, for the choice (\ref{ffi}) $\xi_3=0$ while
$\xi_1$ and $\xi_2$ are real and imaginary parts of the complex
parameter $\eta$, see (\ref{eta}).
The  scalar potential in the theory (\ref{rtact}) comes
from the elimination of $D$ and $F$ terms.
In more transparent notations it reads
$$
V=\frac{g^2}8(| q^A|^2-| \qt_A|^2)^2+
\frac{g^2}2 |\qt_A q^A-\frac{\xi_1}{2}|^2
$$
\beq
\label{n2pot}
+\frac12 (|q^{A}|^2+ |\qt_A|^2)  |a +\sqrt{2}m_A|^2.
\eeq
The QED coupling
constant $g^2$ is small near the quark vacuum in (\ref{rtact}) if
$m\gg\Lambda_2$.

The charge singularity (\ref{acs}) is the root of the Higgs
branch \cite{SW2}. To find it we look for zeros of the
potential in (\ref{rtact}). We have
\begin{equation}
\label{df}
\bar q_{Ap}(\tau_m)^p_f\ q^{fA}\ =\ \xi_m, \quad m=1,2,3.
\end{equation}
(Here $m$ is an adjoint $SU(2)_R$ index, not to be
confused with color indices.)
This equation  determines the Higgs
branch (manifold with $\langle q\rangle\neq0$) which touches the
Coulomb branch at the point (\ref{acs}).
It has non-trivial solutions for $N_f\ge 2$ \cite{SW2}. This is
the reason why we choose $N_f= 2$ for our discussion.

Once quark fields develop non-zero VEV's  on the Higgs
 branch the $U(1)$ gauge
group in (\ref{rtact}) is broken and the photon acquires the mass
\beq
\label{mg}
m^2_{\gamma}=\frac{1}{2}g^2 v^2 ,
\eeq
where we introduce the quark VEV
\beq
\label{qvev}
 |\langle q^{fA}\rangle|^2 =|\langle q^{A}\rangle|^2
+|\langle \tilde{q}_{A}\rangle|^2 = v^2 .
\eeq
As soon as the potential  is zero on the fields
which satisfy constraint (\ref{df}) the moduli
fields which develop VEV's on the Higgs branch are massless.

The number of these massless moduli (dimension of the Higgs branch)
is four. To see this, note that we have eight real scalars subject to
three conditions  (\ref{df}). Also one phase is gauged. Overall
we have 8-3-1=4, which gives us the dimension of the Higgs branch.
Four massless scalars correspond to the lowest components of  one
 short hypermultiplet.
The other quark fields (4 real boson
 states + fermions) acquire  the  mass of  the photon (\ref{mg}).
Together with states from the photon multiplet they form
one long (non-BPS) \ntwo multiplet (cf. \cite{VY}). It has 8 boson
+ 8 fermion states.

If we consider energies much less then the photon mass we can
integrate out heavy scalars and electromagnetic field. Then
we are left with the effective sigma model for light fields living on
the Higgs branch.  The four dimensional Higgs branch is a hyper-Kahler
manifold.  At non-zero $\xi_m$ it has Eguchi-Hanson geometry \cite{EH}.
Like in Sect. 3 we use $SU(2)_R$ rotations to put $\xi_2=0$
so $\eta$ is real, $\eta=\xi_1.$
The convenient parametrization of the metric is as follows \cite{GH}
$$
S_{\sigma} = \int d^4 x \left\{
\left[1-(\frac{\xi_1}{w^2})^2\right]^{-1}(\partial_{\mu} w)^2+
w^2\left[(\partial_{\mu}\theta)^2+
\sin^2{\theta} (\partial_{\mu}\delta)^2\right]
\right.
$$
\beq
\label{ehmet}
\left.
+w^2
\left[1-(\frac{\xi_1}{w^2})^2\right](\partial_{\mu}\gamma+
\cos{\theta} \partial_{\mu}\delta)^2 \right\}.
\eeq
Here the Higgs branch is parametrized by one
modulus field $w$ (which takes values from $\sqrt{\xi_1}$ to
$\infty$)
\beq
\label{w}
w^2=|q^{fA}|^2= |q^{A}|^2 +|\tilde{q}_{A}|^2
\eeq
and three phases $\theta$ ( with values in the interval $(0,\pi)$)
and  $\delta$, $\gamma$ with values in the interval $(0,2\pi)$.

We use  three rotations of broken $SU(2)$ subgroup of
global $SU(N_f=2)\times SU(2)_R$ group to put VEV's of the scalar
fields on the Higgs branch in the form
\beq
\label{ehvev}
<w>=v,\; <\theta>=0,\,<\gamma>=0.
\eeq

One important distinction of the Higgs branch at hand
with the one in \none QED is that the base of the Higgs branch in the
present case is a compact manifold rather then a point.
 The base  of the Higgs branch  is defined as
 a submanifold with the minimal $|q|^2$.
For the case of
Higgs branch  (\ref{1hb})  in \none QED
the base is defined  by the condition
$\tilde{q}=0$ which reduces the Higgs branch to
a single point $\rho=0$.
For the case
of the Higgs branch (\ref{df}) this condition becomes
\beq
\label{qqt}
q^{A}\,=\,\bar{\tilde{q}}^A,
\eeq
which can be obtained from the condition  $\tilde{q}=0$
by $SU(2)_R$ rotation transforming $\xi_3$ into  $\xi_1$, see
\cite{VY}. The condition (\ref{qqt}) reduces the number of
real scalars from eight to four. They are  subject to constraint
(\ref{df}) which boils down to a single condition $2|q^{A}|^2=\xi_1$.
Subtracting $U(1)$ phase we get 4-1-1=2 which is the dimension of the
base of the Higgs branch. Clearly this manifold is a two dimensional
sphere $S_2$. In terms of the coordinates in (\ref{ehmet})
it is parametrized by angles $\theta$ and $\delta$, while
\beq
\label{base}
w=\sqrt{\xi_1},
\eeq
on the base manifold.

\subsection{Flux tubes}

Now let us consider ANO strings  in a generic point on the
Higgs branch in \ntwo QED (\ref{rtact}). The field $a$ is
frozen at its VEV
\beq
\label{afroz}
a=-\sqrt{m}
\eeq
and does not play any role in the string solution.
As we discussed in the previous subsection we have
massive scalars with the mass equal to the mass of the photon
(\ref{mg}) and four massless scalars. Therefore to find
the string solution we use the same method as in Sect. 4.

Namely, our string consist of a BPS core formed by heavy fields and a
tail formed by light fields. To find the BPS core we
impose condition
\beq
\label{qqtvp}
q^{A}\,=\,\bar{\tilde{q}}^A
=\frac1{\sqrt{2}}\vp^A,
\eeq
which reduces the QED (\ref{rtact}) to the Abelian Higgs
model with two complex flavors $\vp^A$ with the potential
\beq
\label{pot2fl}
V= \frac{g^2}{8}\left( |\vp^A|^2-\xi_1\right)^2.
\eeq
Clearly this model possess standard BPS strings \footnote{See however
next section for the discussion on semilocal strings}
with the tension
\beq
\label{bpsten2}
T_{BPS}=2\pi\xi_1
\eeq
which satisfy boundary conditions
\beq
\label{bccore}
w(r\ga R_c)=\sqrt{\xi_1},\; \theta(r\ga R_c)=\theta_0,\;
 \delta(r\ga R_c)=\delta_0,
\eeq
outside the core. Here the size of the core $R_c$ is given by
\beq
\label{coresize}
R_c=\frac{1}{g\sqrt{\xi_1}}.
\eeq

Outside the core    heavy fields
are almost zero and the string is determined by the
classical solution of the one dimensional sigma model (\ref{sigma1d})
(with $\xi_3$ replaced by $\xi_1$) with target space geometry
(\ref{ehmet}). Namely, the tension of the tail is given by
\beq
\label{tail2}
T_{tail}= \frac{2\pi}{\log{L/ R_c}}\,l^2,
\eeq
where $L$ is the length of the string and $l$ is the
length of the geodesic line on the Higgs branch (\ref{ehmet})
between points (\ref{bccore}) and (\ref{ehvev}).

 The initial point of this geodesic line
(\ref{bccore}) is not fixed yet. In principle, it could be any point
on the base submanifold. To fix it we require the string to have
the minimal tension. This boils down to the condition of having
geodesic line of the minimal length between point (\ref{ehvev})
and the point in question on the base submanifold.

 This requirement ensures that  this point has
\beq
\label{bcbase}
w(r\ga R_c)=\sqrt{\xi_1},\; \theta(r\ga R_c)=0.\;
\eeq
Moreover, clearly the phases $\theta$, $\delta$ and $\gamma$ are zero
on the whole geodesic trajectory. We checked this explicitly
solving the Hamilton-Jacobi equation for this geodesic line
following the method of ref. \cite{M}. This means that
the length of the geodesic line is given by
\beq
\label{leh}
l=\int_{\sqrt{\xi_1}}^{v} \frac{dw }{\sqrt{1-(\frac{\xi_1}{w^2})^2}}.
\eeq
With this length the final answer for the tension of the string
becomes
\beq
\label{teneh}
T= 2\pi\xi_1 + \frac{2\pi}{\log{(g\sqrt{\xi_1}\,L)}}\,l^2.
\eeq

If we take the VEV's of scalar fields on the base submanifold
($v=\sqrt{\xi_1}$) the second term in (\ref{teneh}) vanishes and
we get tension of BPS string. Note, that like in Sect. 4 we get
two short BPS  multiplets in this limit from one long
non-BPS string multiplet which exist in a generic point on the
Higgs branch.
In the opposite limit $\xi_1\to 0$ the Eguchi-Hanson manifold becomes
flat and $l\to v$. Besides that the BPS core contribution in
(\ref{teneh}) vanishes and we get (\ref{ct}) obtained
in \cite{Y99} for the case of Higgs branch without FI term
\footnote{Note that the size of electromagnetic core in the limit
$\xi_1\to 0$ is frozen on the value $R_g$, see (\ref{ts})}.

To conclude this section we can reexpress the boundary conditions on
the base manifold (\ref{bcbase}) which serves as a conditions
"at infinity" for the BPS core of the string in terms of the
original quark fields $\vp^A$
of the model with potential (\ref{pot2fl}). Using the standard
relation  of unit vector on $S_2$ in  $O(3)$ sigma model
with quark variables
$$
\cos{\theta}=\frac1{\xi_1}\bar{\vp}_A(\tau^3)^A_B \vp^B,
$$
$$
\sin{\theta}\cos{\delta}=\frac1{\xi_1}\bar{\vp}_A(\tau^1)^A_B \vp^B,
$$
\beq
\label{nfield}
\sin{\theta}\sin{\delta}=\frac1{\xi_1}\bar{\vp}_A(\tau^2)^A_B \vp^B
\eeq
we obtain that point (\ref{bcbase}) on the base
submanifold corresponds to the following values of quark fields
\beq
\label{bpq}
\vp^{A=1}=\sqrt{\xi_1},\; \vp^{A=2}=0.
\eeq
This means that with our choice of scalar VEV's (\ref{ehvev}) the BPS
core is formed only by the first flavor, while the second one remains
unexcited.

\section{Semilocal strings}
\setcounter{equation}{0}

So far our study of flux tubes in \ntwo QED was not complete.
The point is that BPS strings in theories with extended global
symmetry possess an additional zero mode associated with their transverse
size $r_0$. The string parametrized by the size $r_0$ is called
semilocal string (see \cite{AV} for a review). Semilocal string
interpolates between ANO string and 2d sigma model instanton
lifted in four dimensions (lump). At non-zero $r_0$ the semilocal
string has power fall-off of the profile functions at infinity,
instead of the exponential fall-off for ANO string at $r_0=0$.

As we explain
below this leads to a dramatic physical effect - we loose confinement
if a semilocal string is developed instead of ANO string.
In the next subsection we briefly review semilocal strings
and then consider the possibility of their formation
in a generic point on the Higgs branch in \ntwo QCD
with non-zero FI term.

\subsection{BPS semilocal strings}

Let us recall basic features of semilocal strings \cite{AV}.
 The simplest model where they appear is the Abelian Higgs
model with two complex flavors
\begin{equation}
\label{ah2fl}
S_{AH}=\int d^4x\left\{\frac1{4g^2}\,F^2_{\mu\nu}+|\nabla_\mu
\vp^A|^2+\frac{g^2}{8}(|\vp^A|^2-\xi_1)^2\right\}.
\end{equation}
Note, that we already considered this model in Sect. 5.2
discussing the BPS core of the string on the Higgs branch.
 It arises from \ntwo QED (\ref{rtact})
 when we restrict ourselves to the {\em ansatz} (\ref{qqtvp}).

The topological reason for existence of ANO vortices is that
for gauge group $U(1)$ $\pi_1[U(1)]=Z$. On the other hand  we can go to
the low energy limit in (\ref{ah2fl})  restricting ourselves
to the vacuum manifod $|\vp^A|^2=\xi_1$. The vacuum manifold has
dimension 4-1-1=2, where we subtract
one real condition mentioned above as well as  one gauge phase. It
represent two dimensional sphere $S_2$.  Thus, the low energy limit
of theory  (\ref{ah2fl}) is $O(3)$ sigma model.
Now recall that
$\pi_2[S_2]=\pi_1[U(1)]=Z$ and this is a topological reason for
existence of instantons in  two dimensional
$O(3)$ sigma model.  Lifted  in four dimensions they become
a string-like objects (lumps).

So now the question is what is the relation between ANO flux tubes
of QED (\ref{ah2fl}) and lumps of $O(3)$ sigma model.
Clearly the model (\ref{ah2fl}) has ANO strings. Say, if we put the
second flavor to zero this model reduces to the one
flavor model considered in Sect. 4.1. Then ANO string is given
by eqs. (\ref{prof}), (\ref{foe}) for the first flavor while the second
one is zero. However, it turns out (see \cite{H}) that this solution
has zero mode associated with exiting of the second flavor.
This zero mode is parametrized by the parameter $r_0$ which plays the
role of the size of the string in (1,2)-plane.

To find this solution let us modify the standard parametrization
(\ref{prof})  for the ANO string including the second flavor
\begin{eqnarray}
\label{profsl}
\vp^1(x) &=& \phi_1 (r)\, {\rm e}^{i\,\alpha}\ ,\nonumber\\
\vp^2(x) &=& \phi_2 (r)\, ,\nonumber\\
A_i(x) &=&\! \!-2\epsilon_{ij}\,\frac{x_j}{r^2}\ [1-f(r)]\ .
\end{eqnarray}
Note, that the second flavor does not wind at infinity. Therefore,
it has boundary condition $\phi_2 (\infty)=0$ while at $r=0$ it can
be non-zero.

The first order equations for the profile functions here look like
$$
r\frac{\rm d}{{\rm d}r}\,\phi_1 (r)- f(r)\,\phi_1 (r)\ =\ 0\ ,
$$
$$
r\frac{\rm d}{{\rm d}r}\,\phi_2 (r)- (f(r)-1)\,\phi_2 (r)\ =\ 0\ ,
$$
\beq
 -\frac1r\,\frac{\rm d}{{\rm d}r} f(r)+\frac{g^2}{4}\,
\left(\phi_1^2(r)+\phi_2^2(r)-\xi_1\right)\ =\ 0\ .
\label{foesl}
\eeq
The solution to these equations \cite{H,LS} at non-zero $r_0$
is very different from that of ANO string. It has long range power
fall-off at infinity for all profile functions. In particularly,
in the limit of
large transverse size of the string  $r_0\gg 1/g\sqrt{\xi_1}$ it has the
form
$$
\phi_1(r)=\sqrt{\xi_1}\frac{r}{\sqrt{r^2+r_0^2}},
$$
$$
\phi_2(r)=\sqrt{\xi_1}\frac{r_0}{\sqrt{r^2+r_0^2}},
$$
\beq
\label{lump}
f=\frac{r_0^2}{r^2+r_0^2}.
\eeq
This solution has the same tension as ANO string
\beq
\label{lumpten}
T=2\pi\xi_1.
\eeq

We see that scalar fields on the solution (\ref{lump})
belong to the vacuum manifold
$|\vp^A|^2=\xi_1$. This means that we can relate this solution
to the $O(3)$ sigma model lump. To do this we first use
the relation between fields $\vp^A$ and the unit vector
on $S_2$ (\ref{nfield}) and then represent this vector
in terms of complex field $\omega(x)$ via standard relations
$$
\cos{\theta}=\frac{1-|\omega|^2}{1+|\omega|^2},
$$
$$
\sin{\theta}\cos{\delta}=2\frac{Re\,\omega}{1+|\omega|^2},
$$
\beq
\label{o3rel}
\sin{\theta}\sin{\delta}=2\frac{Im\,\omega}{1+|\omega|^2}.
\eeq
With  this substitution the low energy limit of (\ref{ah2fl})
becomes an  $O(3)$ sigma model
\beq
\label{o3}
S_{eff}=\int d^4 x \frac{|\partial_{\mu}\omega|^2}{(1+|\omega|^2)^2}.
\eeq

The standard lump (instanton) solution
 of this model
with center at zero and size $r_0$ looks like
\beq
\label{inst}
\omega_{lump}=\frac{r_0}{x_1+ix_2},
\eeq
where parameter $r_0$ can be made real  by the
shift in the polar angle $\alpha$.

If we now reexpress this solution in terms of quark fields
using relations (\ref{o3rel}), (\ref{nfield}) we arrive at
the solution (\ref{lump}). Thus we identified the semilocal string
in the limit of large $r_0$ (\ref{lump}) with the lump
solution of $O(3)$ sigma model (up to a factorization over $Z_2$).
 The reason for this identification is
that $O(3)$ sigma model is a low energy effective theory for
two-flavor Abelian Higgs model (\ref{ah2fl}).

Now let us come back to our \ntwo QED (\ref{rtact}) and consider
first the vacuum which belongs to the base submanifold
$v=\sqrt{\xi_1}$
of the Higgs branch. This was done in \cite{ADPU,PU}.
In this case the ANO string considered in
the Sect. 5.2 becomes BPS. As we already mentioned
for these VEV's we can look for string solution using the
{\em ansatz} (\ref{qqtvp}) which reduces the bosonic part of our
theory to the two flavor  model ({\ref{ah2fl}). This model has
semilocal strings  so our BPS ANO string is just a particular
case of a semilocal string at $r_0=0$ \cite{ADPU,PU}. Say, in the
opposite limit $r_0\gg 1/g\sqrt{\xi_1}$  the semilocal string is given
by the profile functions (\ref{lump}).

The most physically important consequence of emergence of semilocal
strings is that we loose the monopole confinement. As we already
explained to study confinement we have to consider a string of
a finite length $L$ stretched between heavy monopole and anti-monopole.
Clearly the problem now becomes three dimensional. The string size
$r_0$ does not correspond to a zero mode any longer. Instead, it
is fixed by the separation $L$ at large value $r_0\sim L$. Thus the
monopole flux is not trapped into a narrow flux tube. Instead, it is
spread over a large three dimensional volume  of size of order of
$L$. Clearly this produces a potential between monopole and
anti-monopole with power-like fall-off at large separations,
$V(L)\sim L^{-\gamma}$ ($\gamma>0$). Such potential does not confine.

We see that formation of semilocal strings at a base point on the Higgs
branch lead to a dramatic physical effect -- deconfinement. Therefore
it is quite important to study the possibility of formation of
semilocal strings at a generic point on the Higgs branch. We do it
in the next subsection.

\subsection{Semilocal strings at a generic vacuum on the Higgs branch}

Now we assume that semilocal string can be formed at a generic vacuum on
the Higgs branch of \ntwo QCD  and study the dependence of
its tension on its size $r_0$. As we mentioned in the previous
subsection there is no such dependence for BPS string
in the vacuum which belongs to a   base submanifold,
$r_0$ is associated with the zero mode. We will see below that this is
not the case for a  semilocal string at a
generic vacuum on the Higgs branch.

To study the possibility of formation of semilocal strings it is
sufficient to consider a semilocal string in the limit
of large size  $r_0\gg 1/g\sqrt{\xi_1}$ because in this
limit the semilocal string shows its crucial distinctions from
the ANO one. As we discussed in the previous subsection
in this limit the semilocal string is formed by light fields only.
It does not have core formed by heavy fields and looks like
sigma model lump.

Therefore to study semilocal string in the limit $r_0\gg 1/g\sqrt{\xi_1}$
we can  go to the low energy limit in \ntwo QED (\ref{rtact})
which is given by sigma model (\ref{ehmet})
with Eguchi-Hanson geometry. Assuming that
all fields in this sigma model depend only on coordinates
in $(1,2)$-plane we rewrite the tension of our semilocal string as
follows
$$
T_{semilocal} = \int d^2 x \left\{
\left[1-(\frac{\xi_1}{w^2})^2\right]^{-1}(\partial_{\mu} w)^2+
w^2\left[(\partial_{\mu}\theta)^2+
\sin^2{\theta} (\partial_{\mu}\delta)^2\right]
\right.
$$
\beq
\label{ehsm}
\left.
+w^2
\left[1-(\frac{\xi_1}{w^2})^2\right](\partial_{\mu}\gamma+
\cos{\theta} \partial_{\mu}\delta)^2 \right\}.
\eeq

Now our aim is to find a classical solution of this model
which represents a kind of hybrid of the "tail" solution
of Sect. 5.2 with lump (instanton) solution  of Sect. 6.1.
The boundary conditions of this solution at infinity are given
by the VEV's  (\ref{ehvev})
\beq
\label{ehvevsl}
w(\infty)=v,\; \theta(\infty)=0,\,\gamma(\infty)=0,
\eeq
while the boundary conditions at zero are given by those of the
lump (\ref{lump}), namely $\vp^1(0)=0$ and $\vp^2(0)=\sqrt{\xi_1}$.
Rewriting this  in terms of fields entering the model (\ref{ehsm})
using (\ref{nfield}) we get
\beq
\label{ehzbc}
w(0)=\sqrt{\xi_1},\; \theta(0)=\pi,\,\gamma(0)=0.
\eeq

We see that our solution winds around base cycle of the
Eguchi-Hanson manifold ($S_2$) and then goes away in a non-compact
direction (along $w$). Awaiting for the full solution of the problem
here in this section we consider the case when the VEV of scalar fields
is much larger then the FI parameter, \beq \label{limit} v\, \gg
\,\sqrt{\xi_1}.  \eeq

In this limit the field $w$ goes far away  from the base (at
$w=\sqrt{\xi_1}$) along the non-compact  manifold. When $w\gg \sqrt{\xi_1}$
the geometry in (\ref{ehsm}) becomes flat and the problem reduces to
solving of free equations of motion.  For the radial coordinate
$w$ the solution of the equation $\partial^2 w=0$ reads (cf.
(\ref{scsol}))
\beq
\label{wlog}
w(r)=v\frac{\log{r/r_0}}{\log{L/r_0}},
\eeq
while
\beq
\label{thetagamma}
\theta(x)=0,\;\delta(x)=0,\;\gamma(x)=0
\eeq
like for the "tail" solution of Sect.5.2.
Here we use the lump size $r_0$ as a UV cutoff for the logarithmic
behavior of $w$. This solution is valid at large $r$, $R\ll r \ll L$,
where parameter $R$ will be determined shortly.

Instead for small $r$, $r\la R$ the solution is given by a certain
deformation of the instanton solution of Sect. 6.1 which we denote as
\beq
\label{inst}
(w(x),\theta(x),\delta(x),\gamma(x))= (w_{inst}(x),\theta_{inst}(x),
\delta_{inst}(x),\gamma_{inst}(x)),
\eeq
where $w_{inst}(x)\sim \sqrt{\xi_1}$. Clearly the solution changers its
behavior from (\ref{inst}) to (\ref{wlog}), (\ref{thetagamma})  when
the coordinate $w$ given by the logarithmic  expression  (\ref{wlog})
becomes much larger then $\sqrt{\xi_1}$. This gives
\beq
\label{R}
R = r_0\exp{(\frac{\sqrt{\xi_1}}{v}\log{L/r_0})}.
\eeq

The tension of our semilocal string is now given by the sum
 of tensions of deformed instanton (\ref{inst}) and  the logarithmic
"tail"  (\ref{wlog}), (\ref{thetagamma}),
\beq
\label{slten}
T_{semilocal}=T_{inst}(r_0)+\frac{2\pi v^2}{\log{L/r_0}},
\eeq
where the calculation of the tail contribution is
similar to that in Sect. 2 (see \cite{Y99}).

Of course we do not know the function $T_{inst}(r_0)$ without knowing
the solution (\ref{inst}). What we know is that at $r_0=0$ the semilocal
string reduces to ANO string and deformed instanton (\ref{inst})
goes into BPS core of the string discussed in Sect. 5.2. Thus,
\beq
\label{tinst0}
T_{inst}(r_0=0)=2\pi\xi_1,
\eeq
see (\ref{bpsten2}). Moreover, the central charge of \ntwo algebra
gives us a lower bound for this tension (see, for example, \cite{VY}
for details) \footnote{To prove this formally we can construct
configuration given by  (\ref{inst}) at $r\la R$ and extrapolated by
constant fields at $r\gg R$. Then we  use the SUSY bound for this
configuration.}
\beq
\label{cchbound}
T_{inst}(r_0)\ge 2\pi\xi_1.
\eeq

We see that at least the second term in (\ref{slten}) depend on $r_0$,
so $r_0$ is not associated with conformal zero mode any longer. The
reason for this breaking of conformal invariance at the classical level
is the non-compactness of the Higgs branch.
To see this note, that in the last subsection we
were dealing with lump solution which maps two dimensional space onto
compact base submanifold $S_2$ of the Eguchi-Hanson Higgs branch.
The tension of this lump solution is a constant (\ref{lumpten})
determined by the volume of $S_2$. The collective coordinate
$r_0$ in this case is associated with the conformal zero mode.

As we move to a generic vacuum on
the Higgs branch our semilocal string solution becomes a map onto
a non-compact manifold. In order to get a finite string tension we
use a cutoff in the target space introduced by the VEV of the
scalar fields $v$. This cutoff in the target space requires an IR cutoff
$L$ in the two dimensional $(1,2)$-plane.  We already discussed how
this works in Sect.2. The logarithmic solution to the free equations
of motion (\ref{wlog}) cannot go to its VEV at infinity. In order to
insure that the scalar fields reach their VEV's at infinity we have to
introduce a IR cutoff, say small masses for light fields (like in
Sect. 2,4) or the finite length of the string $L$. Clearly this
breaks the conformal invariance and produces the dependence of
the string tension (\ref{slten}) on $r_0$.

Now minimizing the string tension (\ref{slten}) with respect to $r_0$
using (\ref{tinst0}) and (\ref{cchbound}) we find that
\beq
r_0=0.
\eeq
In the limit $r_0\to 0$ the
semilocal string reduces to ANO string considered in Sect. 5.2.
In particular, the string tension is given by  (\ref{teneh})
in the limit $l\to v$.
Note, that quantum fluctuations of $r_0$ are negligibly small,
suppressed  by the world sheet volume of the string.

 This means that  semilocal strings are not  formed in the
generic vacuum on the Higgs branch at least in the limit $v\gg
\sqrt{\xi_1}$.  This conclusion is in accordance with the general
expectation \cite{AV} that semilocal strings are not stable in type I
superconductor and reduce to ANO strings.

\subsection{Deconfinement phase transition}

The conclusion
made in the previous subsection
that semilocal strings are not developed at a generic
vacuum on the Higgs branch (at least if $v\gg \sqrt{\xi_1}$ ) and
we are dealing with ANO strings has rather important physical
consequences. This means that we have a confinement phase for
monopoles in these vacua with confining potential determined
by the ANO string tension (\ref{teneh}), see (\ref{cpotcurved}).
Instead, for the vacua on the base submanifold $v=\sqrt{\xi_1}$
semilocal strings are formed and we have deconfinement phase.

Therefore, it is clear that we can expect   confinement/deconfinement
phase transition at certain intermediate critical value $v_c$. To find
this critical value we have to study semilocal string solution
for vacua which are
not far away from the base submanifold, $v\sim \sqrt{\xi_1}$. This is
left for a future work.  Here we put forward a plausible
conjecture that in fact
\beq
\label{conj}
v_c= \sqrt{\xi_1},
\eeq
which means that we have confinement phase for all vacua on the Higgs
branch except the base submanifold.  The reason for this conjecture
is that nothing special happens at any intermediate value $v_c$.
Instead, the value $v_c= \sqrt{\xi_1}$   is physically distinct
because at this value of scalar VEV strings become BPS-saturated
(one long non-BPS string multiplet becomes two short BPS multiplets,
see Sect. 5.2).

This perfectly matches results of \cite{Y99} where
unbroken \ntwo QCD with two degenerative flavors at $\xi_1=0$ was
considered. At zero $\xi_1$ the geometry of Higgs branch becomes flat,
given by $R^4/Z_2$ and the $S_2$ cycle shrinks to a singular point at
the origin.
It was shown in \cite{Y99}  that monopoles are in the confinement phase
on the Higgs branch at any non-zero $v$. As $v$ goes to zero
the transverse size of ANO string becomes infinite while its tension
goes to zero,
see (\ref{ts}), (\ref{ct}), so monopole confinement becomes
unobservable.  Introducing the  non-zero FI parameter $\xi_1$ we
resolve the
singularity at the origin   and see that at the base cycle $S_2$
we have deconfinement phase while on the rest of the Higgs branch we
have monopole confinement.

To conclude this section we would like to stress that the above
discussion of confinement/deconfinement phase transition refers  only
to \ntwo QED (\ref{rtact}) considered on its own right.
In contrast, if we start from non-Abelian  softly broken \ntwo
$SU(2)$ QCD the story is rather different.  The point is that
( as we mentioned in Sect. 5.1) the \ntwo QED (\ref{rtact})
is the low energy description
  of the underlying non-Abelian theory only in the
special limit
\beq
\label{ntwolimit}
\mu\to 0,\; m\to\infty,\; \xi_1=const,
\eeq
which ensures that \ntwo supersymmetry is preserved. In this case
the superconductivity on the base submanifold is on the
border between type I and type II and we have BPS strings
which eventually become semilocal strings in the presence of
$SU(2)$ global flavor symmetry.

Generically if we take into account corrections in $\mu/m$
the \ntwo supersymmetry is broken down to \none.
As it is shown in \cite{VY} superconductivity becomes of type I.
For the type I superconductor semilocal strings become
unstable \cite{AV}. In particular, we expect that the string
tension even for the string in the vacuum on the base
submanifold becomes a function of $r_0$
\beq
\label{1lump}
T_{lump}\sim 2\pi\xi_1\left(1 + const\; r_0^2(\delta m)^2 +\cdots
\right),
\eeq
where $\delta m$ is the splitting of masses in the former \ntwo
multiplet. This splitting is given by \cite{VY}
\beq
\label{split}
\delta m \sim \mu.
\eeq
We conclude that in this case the semilocal string is not
developed. Minimization of (\ref{1lump}) with respect to $r_0$
gives
\beq
\label{r0}
r_0\,=\, 0.
\eeq
Note again that quantum fluctuation in $r_0$ are suppressed
by the volume of the string world sheet.

This means that we do not have deconfinement phase even on the base
submanifold of the Higgs branch in QCD with \ntwo supersymmetry
softly broken down to \none. All vacua
on the Higgs branch of this theory give rise to the confinement phase
for monopoles.

\section{Conclusions}

In this paper we studied flux tubes on Higgs branches in supersymmetric
gauge theories. Our main conclusion is that although Higgs branches
ensure presence of massless scalars flux tubes still exist
and give rise to a confining potential. However due to the presence of
massless scalars the string tension is no longer a constant. It becomes
a slow (logarithmic) function  of the length of the string $L$. Still
it produces
a confining potential of type (\ref{cpotcurved}).

First we studied \none QED with FI $D$-term. This theory has a two
dimensional Higgs branch. The solution for the string in a generic
point on this Higgs branch is given by "BPS core" formed by heavy
fields and a "tail" formed by light fields living on the Higgs branch.
Finding the solution for the "tail" reduces to finding a classical
solutions of the sigma model on the Higgs branch. The tension
of the string is given by the
sum of tensions of "BPS core" and
the "tail", see (\ref{tenL}).

Next we studied \ntwo QCD  with gauge group $SU(2)$ and two
flavors of quarks with the same mass. This theory has a four
dimensional Higgs branch. We deformed this theory with the  mass term
for the adjoint field. To the leading order in the  deformation
parameter $\mu$ \ntwo supersymmetry is not broken in the
 effective QED describing the low energy limit of this theory.
Higgs branch has an Eguchi-Hanson geometry.

We showed that far away from the base cycle $S_2$ of the Higgs branch
stable ANO strings are formed. Their tension is again given by the
sum of tensions of "BPS core" and
the "tail", see (\ref{teneh}). These strings produce confinement of
monopoles with the confining potential of type (\ref{cpotcurved}).
However, in the vacua on the base cycle $S_2$
of the Higgs branch the story is rather
different. Semilocal BPS strings are developed in vacua on the base
submanifold  leading to a deconfinement regime. We conjecture that the
confinement/deconfinement phase transition occurs exactly
at the base cycle of the Eguchi-Hanson space.

If we introduce breaking of \ntwo supersymmetry down to \none in  this
theory  it turns out that the deconfinement phase disappears and we
have monopole confinement on the whole Higgs branch.

\section*{Acknowledgments}

We are grateful to Alexander Gorsky, Mikhail Shifman,
David Tong and
Arkady Vainshtein for  helpful discussions.
This work is supported in part by INTAS grant No.~00-00334.
The work of
A.~Y. is   also supported  by the Russian Foundation for Basic
Research   grant No.~02-02-17115
and by Theoretical Physics Institute
at the University of Minnesota.

\end{document}